# Partial Times Delay in Elastic Electron Scattering by Rectangular Potential Well with Arising Discrete Levels

M. Ya. Amusia[1,2], A. S. Baltenkov[3]
and
I. Woiciechowski[4]

[1] Racah Institute of Physics, the Hebrew University, Jerusalem 91904, Israel
[2] Ioffe Physical-Technical Institute, St. Petersburg 194021, Russia
[3] Arifov Institute of Ion-Plasma and Laser Technologies,
100125, Tashkent, Uzbekistan
[4] Alderson Broaddus University, WV 26416, Philippi, USA

**Abstract**

We have studied the times delay of slow electrons scattered by a spherically symmetric rectangular potential well as functions of the well parameters. We have shown that the electron interaction with the scattering center qualitatively depends on the presence of discrete levels in the well. While electron retention dominates for the potential well with no discrete levels, the appearance of a level leads to the opposite situation where the incident electron hardly enters the scatterer. Such a behavior of the time delay is universal since we found it not only for the first *s*-level but also for the following arising *s*-, *p*-, and *d*-levels.

**1. Introduction**

The earlier experiments with atomic and molecular photoionization under attosecond laser pulses have resulted in the first observation of the time delay in photoemission [1-4]. Nowadays, the attosecond chronoscopy is a rapidly developing field of research (see, for example, [5-7] and references therein). Photoionization appeared to be not an instantaneous process, contrary to the initial believe. Departure of the photoelectron wave packet delays relatively to the arrival of the photon pulse. The typical time delay is a few attoseconds. This physical effect bears the name "time delay in *half-scattering process*" in contrast to the time delay originally introduced by Eisenbud, Wigner and Smith [8-10]. Their time delay, referred to as the EWS-time delay, is due to the particle capture and retain for some time by a short-range potential in the resonance *elastic scattering process*. In the half-scattering process, the photoelectron moves from the center of the ionized system to its border, while in the elastic scattering process, the classical trajectory of particle crosses the system from the border to border, passing through the center. For this reason, the time delay in the half-scattering process is as twice as less than the EWS-time delay. In the special case of scattering by potential of spherical symmetry, the EWS-time delay $\delta_l(E)$ is simply the derivative of the particle scattering phase shift $\delta_l(E)$ with respect to the particle kinetic energy $E$ [6]

$$\tau_l(E) = 2\hbar \frac{d\delta_l(E)}{dE} \qquad (1)$$

The experimental observations of the times delay in the half-scattering *photoionization process* [11] show that the electron travel times through the scattering region are significantly different for the partial electron wave packets with different orbital momenta *l*. We should expect similar situation in the particle *elastic scattering processes*. There is, therefore, a necessity of a comparative study of the partial components of the classical EWS-times delay. Such a study is the goal of the present paper.

In our recent paper [12], we investigate the partial EWS-times delay in the process of slow elastic electron scattering by the Dirac bubble potential [13]. This potential is a good model for the fullerene $C_{60}$ shell. We found that the sign of the partial time delay depends upon presence of a discrete level with the corresponding orbital moment *l* in the potential well. Namely, the $l^{\text{th}}$ time delay is positive when there is no discrete *l*-level in the well, and it becomes negative when such a level appears. One can expect similar behavior of the time delay if not only the first, but also the second-, third-, and so on discrete levels appear. This hypothesis cannot be verified for the Dirac bubble potential since it has only one discrete level in the potential well with the given orbital moment *l*.

To overcome this limitation, we have investigated the EWS-time delay behavior in the case of slow electron elastic *s*-scattering by the spherically symmetric square-potential well $V(r)$ [14]. Such a potential, in contrast to the Dirac bubble potential is capable to support any numbers of *s*-discrete levels. Specifically, we studied the EWS-times delay as functions of the well parameters, such as the potential well radius *R*, and depth *U*. We showed that the duration of electron interaction with the scattering center qualitatively depends on presence of discrete levels in the well. While electron retention dominates for the potential well with no discrete levels, the appearance of a level leads to the opposite situation where the incident electron hardly enters the scatterer. Such a behavior of the time delay is universal since it takes place not only for the first *s*-level but also for the following *s*-levels appearing in the potential well.

The goal of the present paper is twofold. First, we generalize the results of paper [14] to the case of nonzero orbital moment *l* of the electron partial wave. Second, we compare the *p*- and *d*-EWS-times delay with the *s*-times delay. The dependence of the scattering phase shifts $\delta_l(k)$ and, therefore, the dependence of the EWS-times delay $\tau_l(k)$ upon the type of function $V(r)$ can be in either hidden or explicit forms. Choosing $V(r)$ as the spherically symmetric square well potential with radius *R* and depth *U*, we acquire the possibility to derive the desired dependences in an explicit form, complementing the analytic approach by some numerical calculations.

The organization of the paper is as follows. Sections 2 contains the main general formulas for the derivatives of the *s*- and *p*-phase shifts in the rectangular potential well. Section 3 presents the general equations for the potential well parameters, for which the *s*-, *p*- and *d*-discrete levels appear. Section 4 presents numerical calculations of the derivatives of the phase shifts and the corresponding EWS-times delay. Subsection 4.1, includes calculations of the phase shifts and times delay as functions of the potential well depth *U*. Subsection 4.2 gives these quantities as functions of the electron momentum *k* . Subsection 4.3 presents the *R*-dependence of derivatives of the phase shifts and the corresponding EWS-time delays. Conclusions are given in Section 5. Appendix collects the details of our consideration for *d*-partial electron wave.

## 2. Main formulas

Let an electron with the wave vector $k$ move in continuum above the attractive potential well

$$V(r) = \begin{cases} -U, & r < R, \\ 0, & r \geq R. \end{cases} \quad (2)$$

The following formula connects the electron wave vectors inside the potential well $q$ and outside the well $k$:

$$\frac{k^2}{2} + U = \frac{q^2}{2} \quad (3)$$

Here and throughout the paper, we employ the atomic system of units. Inside the potential well, the radial part of the electron wave function is the regular spherical Bessel function [15]

$$R_{kl}(r) = A j_l(qr), \quad r < R \quad (4)$$

where $A$ is a constant. Beyond the range of $V(r)$, we always take $R_{kl}(r)$ to be in the form of a linear combination of the regular and irregular spherical Bessel functions, $j_l(x)$ and $n_l(x)$, respectively

$$R_{kl}(r) = j_l(kr)\cos\delta_l - n_l(kr)\sin\delta_l \quad \text{at } r \geq R, \quad (5)$$

with the following asymptotic behavior

$$j_l(x \to \infty) = \sin(x - \pi l/2)/x; \quad n_l(x \to \infty) = -\cos(x - \pi l/2)/x. \quad (6)$$

The phase shifts $\delta_l(k)$ in Eq. (5) are calculated by matching the logarithmic derivatives of the electron wave functions inside and outside the potential well (Eq. (4) and Eq. (5)) at the point $r = R$. The matching condition from the outside results in the expression [16]

$$k \frac{j_l'(kR)\cos\delta_l - n_l'(kR)\sin\delta_l}{j_l(kR)\cos\delta_l - n_l(kR)\sin\delta_l} = \gamma_l, \quad (7)$$

while the logarithmic derivative of the "inner" wave function is as follows

$$\gamma_l = q \frac{j_l'(qR)}{j_l(qR)}. \quad (8)$$

Prime symbol at the functions $j'_l(\rho)$ and $n'_l(\rho)$ in Eqs. (7) and (8) denotes differentiation with respect to $\rho$. Solving Eq. (7) for $\delta_l(k)$ we obtain the formula for tangent of the phase shift [16]

$$\tan\delta_l = \frac{kj'_l(kR) - \gamma_l j_l(kR)}{kn'_l(kR) - \gamma_l n_l(kR)}. \qquad (9)$$

For *s*- and *p*-states, the matching conditions Eqs. (7) and (8) can be written in the following explicit forms:

$$k\cot(kR + \delta_0) = q\cot qR \qquad (10)$$

for the *s*-wave, and

$$\frac{1}{k^2}[1 - kR\cot(kR + \delta_1)] = \frac{1}{q^2}[1 - qR\cot qR] \qquad (11)$$

for the *p*-partial electron wave. In Eqs. (10) and (11), the *s*- and *p*-phase shifts $\delta_0(k)$ and $\delta_1(k)$ calculated with Eq. (9), are given in the explicit form in Appendix (see Eq. (A15)). Applying operator $\partial/\partial k$ (denoted by a dot) to both sides of Eqs. (10) and (11) we obtain the following expressions for derivatives of the phase shifts with respect to the wave vector $k$

$$\dot\delta_0 = \frac{\partial \delta_0}{\partial k} = -R + \frac{\sin 2(kR + \delta_0)}{2k} + W_0(x,y) \qquad (12)$$

$$\dot\delta_1 = \frac{\partial \delta_1}{\partial k} = -R - \frac{\sin 2(kR + \delta_1)}{2k} + \frac{2\sin^2(kR + \delta_1)}{k^2 R} + W_1(x,y) \qquad (13)$$

where the functions $W_0(x,y)$ and $W_1(x,y)$ are as follows

$$W_0(x,y) = R\left(1 - \frac{\sin 2x}{2x}\right)\left[\frac{\sin(y + \delta_0)}{\sin x}\right]^2, \qquad (14)$$

$$W_1(x,y) = R\left[x^2 + \left(\frac{x}{2} - \tan x\right)\sin 2x\right]\left[\frac{\sin(y+\delta_1) - y\cos(y+\delta_1)}{y(\sin x - x\cos x)}\right]^2 \qquad (15)$$

Here, the dimensionless variables $x = qR$ and $y = kR$. Comparing Eqs. (12) and (13) with the lower limits of the phase shift derivatives given in paper [9]

$$\dot{\delta}_0 > -R + \frac{\sin 2(kR+\delta_0)}{2k}, \qquad (16)$$

$$\dot{\delta}_1 > -R - \frac{\sin 2(kR+\delta_1)}{2k} + \frac{2\sin^2(kR+\delta_1)}{k^2 R}, \qquad (17)$$

we conclude that the functions $W_0(x,y)$ and $W_1(x,y)$ are always positive. Eqs. (16) and (17) determine the asymptotic behavior of the derivatives when $k \to 0$. According to these inequalities, there are general restriction on the phase shift derivatives with any orbital momenta $l$ that follows from the principle of causality. The principle states, "The scattered wave cannot leave the scatterer before the incident wave has reached it" [9]. Below we will use the Wigner's formulas (16) and (17) in the numerical calculations.

For higher orbital momenta $l \geq 2$, the matching condition leads to expressions that are too complicated to calculate the derivative directly. For $l \geq 2$, it is better to use another method for $\dot{\delta}_l(k)$ calculation. We present the details of the method in Appendix along with its application to the $d$-phase shifts.

The following relations connect the partial EWS-times delays with $\dot{\delta}_0$ and $\dot{\delta}_1$

$$\tau_l(k,U,R) = 2\frac{d\delta_l}{dE} = 2\frac{d\delta_l}{dk}\frac{dk}{dE} = \frac{2}{k}\dot{\delta}_l(k,U,R), \qquad (18)$$

where both the scattering phase derivatives and EWS-times delay are functions of three variables $k$, $U$, and $R$. For small electron energies $E = k^2/2$, according to the Wigner threshold law [17], the phase shifts obey to the condition $\delta_l(k \to 0) \sim k^{2l+1}$. The EWS-times delay near the threshold, therefore, have the asymptotic behavior

$$\tau_l(k,U,R)_{k \to 0} \propto \pm k^{2l-1}. \qquad (19)$$

The $s$-partial EWS-time delay in this limit goes to infinity as $\tau_s(k,U,R)_{k \to 0} \propto \pm 1/k$. Below, we consider the function $\tau_s(k,U,R)$ for small but finite values of the electron momentum $k$. For higher orbital moments $l>0$, the partial EWS-time delay at the threshold equals to zero.

**3. The discrete level arising in the rectangular potential well**

Paper [14] demonstrates that the partial EWS-times delay have specific peculiarities when new discrete levels in the potential well arise. Let us determine the parameters of the potential well with new arising levels. For discrete states with the binding energy $I = -\kappa^2/2$ arising in the rectangular potential well given by Eq. (2), the electron wave vectors inside and beyond the well ($q$ and $\kappa$, respectively) are related as

$$U - \frac{\kappa^2}{2} = \frac{q^2}{2} \qquad (20)$$

The solutions of wave equation that decrease for large values of $r$ as $e^{-r}$ are the spherical Hankel functions of the first kind [15]

$$R_{\kappa l}(r) = B_l h_l^{(1)}(i\kappa r) = B_l [j_l(i\kappa r) + i n_l(i\kappa r)], \quad r > R. \tag{21}$$

Both the condition of matching the logarithmic derivatives of the wave functions Eq. (4) and Eq. (21) at the point $r=R$ and the asymptotic behavior of the wave function at $\kappa \to 0$ result in the equation for the potential well parameters when the discrete $s$-level with zero binding energy $\kappa^2/2 = 0$ appears in the well

$$\cot qR = 0. \tag{22}$$

The potential well depths $U_s^{(n)}$ are related with the potential well radius when the $n^{th}$ consecutive level of $s$-symmetry just appears in the well. The following equation gives this relation

$$U_s^{(n)} = \frac{(2n-1)^2 \pi^2}{8R^2}. \tag{23}$$

Performing the same calculations for $p$-levels, we obtain

$$j_0(qR) = 0 \tag{24}$$

For the parameters of the potential well when $n^{th}$ consecutive level of $p$-symmetry just appears in the well, we obtain

$$U_p^{(n)} = \frac{n^2 \pi^2}{2R^2}. \tag{25}$$

It is straightforward to generalize Eqs. (22) and (24) for states with higher orbital moments $l \geq 2$. The general condition for the level with the $l$-symmetry arising in the well reads [16]

$$j_{l-1}(qR) = 0. \tag{26}$$

**4. Numerical calculations**
In this section, we first numerically investigate the behavior of the $s$- and $p$-partial EWS-times delay $\tau_l(U)$. Then, we study the behavior of the $s$- and $p$-phase shift derivatives $\dot{\delta}_l(k)$ and the partial EWS-times delay $\tau_l(k)$ as functions of the electron wave vector $k$. Finally, we investigate the behavior of the $s$- and $p$-partial EWS-times delay $\tau_l(R)$ versus the potential well radius $R$.

## 4.1. *U*-dependence

In the computations performed, we assume that the potential well depth $U$ in Eqs (12) and (13) is a variable while the radius of the potential $R$ is a constant equal for certainty to 2 at. un. Figure 1 presents the EWS-times delay $\tau_s(U)$ and $\tau_p(U)$ as functions of $U$ for several fixed values of the electron wave vectors $k$. The case of extremely small electron energies $k=0.1$ and $k=0.2$ is represented in two upper panels.

For the potential wells with no bound states (the intervals $0 < U < U_s^{(1)}$ or $0 < U < U_p^{(1)}$), the time delay is positive and it switches its sign after the first discrete *s*- or *p*-level appears in the well when $U$ is increasing. As a result, small changes in $U$ in the vicinities of the first critical values $U_s^{(1)}$ and $U_p^{(1)}$ lead to sudden jumps of the functions $\tau_s(U)$ and $\tau_p(U)$. This specific feature in the behavior of the EWS-time delay for small electron energies is universal since it exists for all other critical values of the variable $U$. Note that the amplitudes of the peaks of the curves presented in two upper panels of Figure 1 decrease when the $k$ values increase. For example, when the wave number $k$ increases by a factor of two, from $k = 0.1$ to $k = 0.2$, the peak value for the *s*-wave at the first calculated point after the critical value $U_s^{(1)}$ decreases by about a factor of nine. For *p*-wave, the difference between the same peaks is only about twice as large.

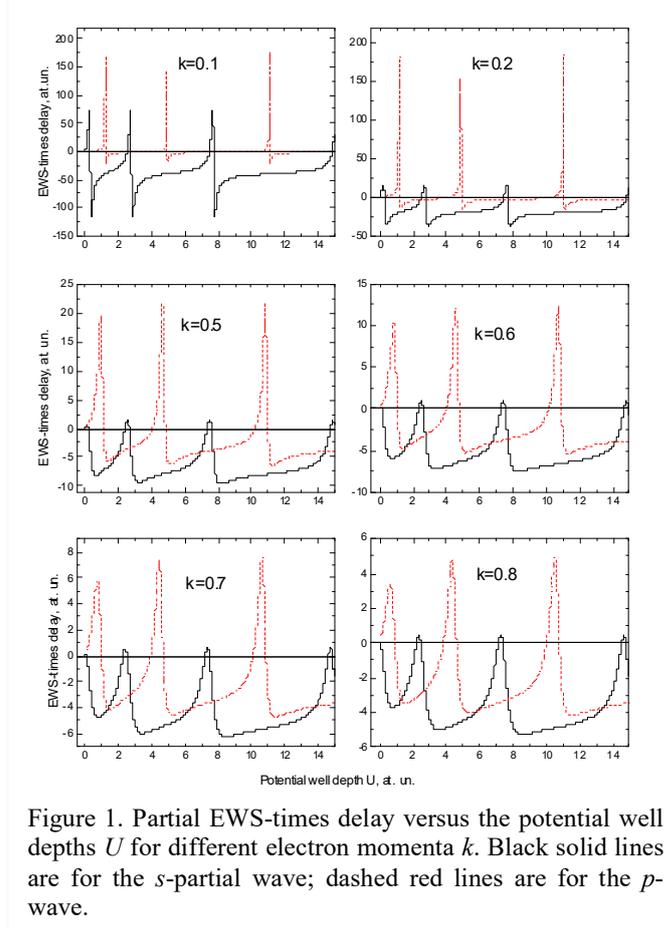

Figure 1. Partial EWS-times delay versus the potential well depths $U$ for different electron momenta $k$. Black solid lines are for the *s*-partial wave; dashed red lines are for the *p*-wave.

The relatively simple resonance behavior of the functions $\tau_s(U)$ and $\tau_p(U)$ for very small wave numbers $k$ changes for larger $k$ values. It is illustrated by four lower panels in Figure 1, which present the calculated functions for the EWS-times delay at $k = 0.5, 0.6, 0.7, 0.8$. We observe here a much more complex picture. The positions of the peaks shifts from the values given by Eqs. (23) and (25). Table 1 gives the shifts of the three zero positions on the $U$ axes.

Table 1. The positions of the first three resonances in
a) *s*-EWS-time delays $\tau_s(U)$

| $k$ | Eq. (23) | 0.1 | 0.2 | 0.5 | 0.6 | 0.7 | 0.8 |
|---|---|---|---|---|---|---|---|

| $U_s^{(1)}$ | 0.308 | 0.307 | 0.285 | 0.195 | 0.135 | 0.075 | 0.015 |
| --- | --- | --- | --- | --- | --- | --- | --- |
| $U_s^{(2)}$ | 2.776 | 2.775 | 2.745 | 2.655 | 2.595 | 2.535 | 2.445 |
| $U_s^{(3)}$ | 7.711 | 7.695 | 7.695 | 7.575 | 7.545 | 7.455 | 7.395 |

b) $p$-EWS-time delays $\tau_p(U)$

| $k$ | Eq. (25) | 0.1 | 0.2 | 0.5 | 0.6 | 0.7 | 0.8 |
| --- | --- | --- | --- | --- | --- | --- | --- |
| $U_p^{(1)}$ | 1.234 | 1.225 | 1.220 | 1.105 | 1.045 | 0.985 | 0.925 |
| $U_p^{(2)}$ | 4.935 | 4.915 | 4.910 | 4.795 | 4.765 | 4.675 | 4.615 |
| $U_p^{(3)}$ | 11.103 | 11.095 | 11.090 | 10.975 | 10.915 | 10.855 | 10.795 |

Note the very different scales of oscillations of the EWS-times delay in Figure 1. In the first and second panels, the range of the time variations is from about –100 to +100 at. un., while in the lower right panel this range is only between –6 and +6 at. un. This is an illustration of the following general behavior of the phase shift derivative [18]

$$\dot{\delta}_l(k \to \infty) \to 0. \quad (27)$$

### 4.2. $k$-dependence

In the computations performed, we assume that both the potential well depth $U$ and the potential well radius $R=2$ at. un. are constants, while the wave vector $k$ is a variable. Let us begin from the phase shift derivatives. In Figure 2, the left panel presents the $s$-phase shift derivatives $\dot{\delta}_0(k)$ as functions of $k$. The right panel presents the functions $\dot{\delta}_1(k)$. In the upper left graph, we have two pairs of curves calculated using Eq. (12) and two pairs of curves obtained using Eq. (16). We choose two specific values of potential well depth $U$: one is $U = 0.2$ at. un. that is below $U_s^{(1)} = 0.308$ at. un. and the other one, $U = 0.4$ at. un., that is above $U_s^{(1)}$. The fist value corresponds to the case with no discrete levels in the well. The second value is for the case when the potential well has the first $s$-level. The pairwise consideration of these well depths allows us to illustrate the principally different behavior of the *s*-phase shift derivative before and after the *s*-discrete level appearance. In the limit of $k \to 0$, the functions $\dot{\delta}_0(k)$ are approaching constant values, but the constants are of opposite signs. For the potential well with no *s*-level, the constant is positive, while it is negative when the discrete level appears. We therefore can conclude that the derivative of the phase shift switches the sign when the first discrete level arises in the well.

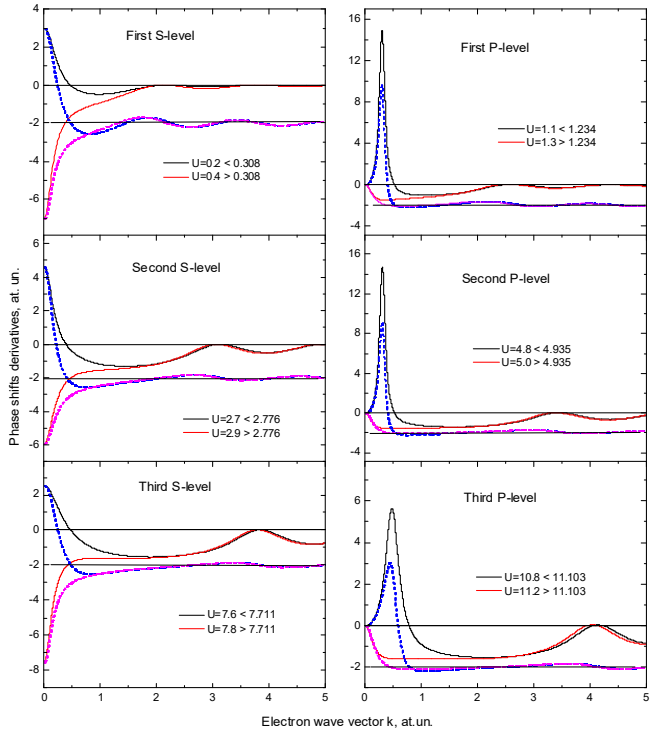

Figure 2. The $s$- and $p$-phase shift derivatives as functions of the electron wave vector $k$. Left panel is the $s$-phase shift; right one is the $p$-phase shift. Solid black and solid red lines are for the phase shifts for potential wells with depths before and after level arising, respectively. The critical values of the first three levels in the wells are: $U_s^{(n)}$=0.308, 2.776, and 7.711 from Eq. (23) for the $s$-state and $U_p^{(n)}$=1.234, 4.935, and 11.103 from Eq. (25) for the $p$-state. Dashed blue and magenta lines present the calculations with the Wigner formula Eqs. (16) and (17), correspondingly.

With increasing *k*, the derivative follows the asymptotic behavior given by Eq. (27). The *k*-dependencies calculated with the Wigner formula Eq. (16) for small electron momenta $k \to 0$ coincide with the *k*-dependencies calculated with Eq. (12), but with one difference. When *k* is increasing the derivative approach a different limit, $\dot{\delta}_0(k) \to -R = -2$. One can see in the Figure the violation of the general restriction $\dot{\delta}_0(k) + R > 0$ (see text after Eq. (17)) at small electron energies. According to Wigner, this violation is the manifestation of the scattering particle wave nature. The restriction $\delta'_0(k) + R > 0$ for large *k* is essentially preserved [9]. We can observe the same picture in the vicinities of the second and third discrete levels (second and third graphs in the left panel).

The right panel of Figure 2 presents the derivatives $\dot{\delta}_1(k)$ calculated with Eqs. (13) and (17). Similar to the *s*-scattering, we have two families of the curves. They correspond to the potential well without bound states and the potential well with the first *p*-bound state arisen (upper right graph). The second and third right graphs, respectively, present the results of the calculations of the phase shift derivatives for the well depths around the second and third *p*-states arisen in the well. In contrast to the functions $\dot{\delta}_0(k)$, one has $\dot{\delta}_1(k \to 0) = 0$ that increases with electron momentum *k* growth as $k^2$. The positions of the $\dot{\delta}_1(k)$ peaks slightly shift to larger *k* with the number of the *p*-level increase. The general behavior of the curves in the left and right panels is similar. Appearance of the discrete *p*-levels in the well leads to changing the sign of the function $\dot{\delta}_1(k)$.

Since the *s*-EWS-times delay near the threshold approaches infinity as $\tau_0(k) \propto \pm 1/k$ (see Eq. (19)), we can compare the times delay for very small but finite values of the electron momentum (*k*=0.01 at. un. was chosen for certainty). For the first pair of the curves (upper left graph in Figure 2), the jump of the EWS-time delay in the vicinity of $U_s^{(1)}$ is $\Delta \tau_0 \approx 2 \cdot 10^3$ at. un. (from $+0.6 \cdot 10^3$ at. un. to $-1.4 \cdot 10^3$ at. un.). For the second pair of the curves (middle left graph), the jump of the time delay is also $\Delta \tau_0 \approx 2 \cdot 10^3$ at. un. For the last pair of derivatives in the lower left graph, we obtain again the same value of the jump (from $+0.5 \cdot 10^3$ at. un. to $-1.5 \cdot 10^3$ at. un.). The *k*-range, where those jumps are observable, is a very narrow interval of electron momenta near the zero *k* value. For this reason, the graphical representation of the function $\tau_0(k)$ is not enough informative.

The situation is different for *p*-electron waves. Figure 3 presents the *p*-EWS-times delay as functions of *k* calculated using Eq. (18) with the *p*-phase shift derivatives $\dot{\delta}_1(k)$ evaluated at the previous step (right panel of Fig. 2). In all the cases, $\tau_1(k)$ is positive before the level arising, switching to negative values when the level appears. The peaks of the curves in Figure 3 slightly shift to larger $k$ with the consecutive number $n$ of the $p$-level while the maxima of the curves $\tau_1(k)$ decrease with the growth of $n$.

**4.3. R-dependence**

In the calculations performed here, we assume that the potential well radius $R$ is a variable while the potential well depth $U$ is a constant value. The relationships for the critical radii, for which $s$- and $p$-discrete levels in the potential well appear, read (see Eqs. (22) and (24)):

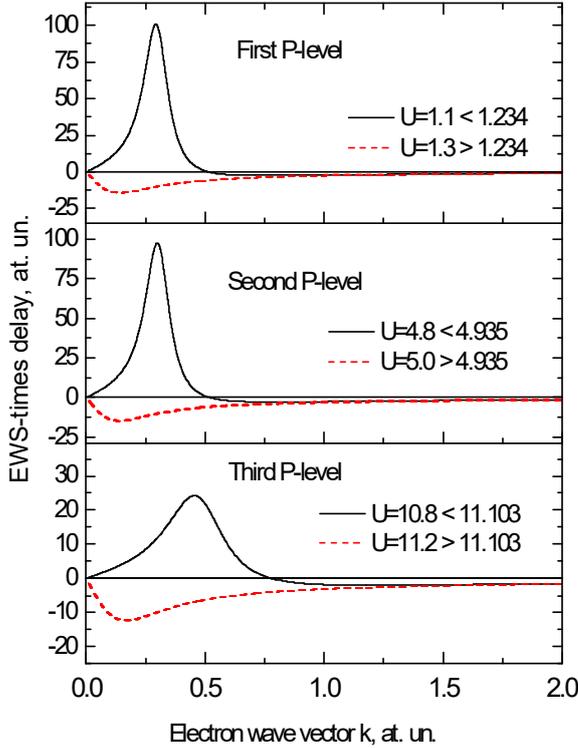

$$R_s^{(n)} = \frac{(2n-1)\pi}{2\sqrt{2U}}; \qquad R_p^{(n)} = \frac{\pi n}{\sqrt{2U}}. \tag{28}$$

Here $n$, just as above, denotes the consecutive numbers of the $s$- or $p$-level in the well. Figure 4 depicts the derivatives of the phase shifts $\dot{\delta}_0(R)$ and $\dot{\delta}_1(R)$ and the EWS-time delays $\tau_0(R)$ and $\tau_1(R)$ as functions of $R$ for fixed both electron wave numbers $k$ and potential depths $U$. The left panel of this figure presents the results for $s$-partial waves, while the right panel presents the results for $p$-waves. Upper graphs depict the sums $\dot{\delta}_l(R) + R$. Causality principle dictates general restriction $\dot{\delta}_l(R) + R > 0$ (see Eq. (2)

Figure 3. The EWS-times delay $\tau_1(k)$ as functions of the electron wave vector $k$ in the vicinities of the first three $p$-levels arising in the potential well. Solid black lines are for the depth before level arising; dashed red lines are for the depth after $p$-level arising.

in [9]). Therefore, the curves in this graph should not cross the $R$ axis since the scattering center $V(r)$ has a finite radius. Thus, the causality principle is applicable. On the other hand, according to Eq. (9) the wave nature of particles does permit some infringement of this restriction. Indeed, we can locate the small negative segment in the interval of $R$ approximately between 6 and 9 at. un. The values become all positive for larger $R$.

According to the lower graphs of Figure 4, times delay $\tau_0(R)$ and $\tau_1(R)$ are alternating functions of $R$. The zeroes of these functions are close to the critical values of the potential well radii. Table 2 where the corresponding values are collected illustrates this fact. Using Eq. (28), we calculate the critical values of the first three ($n$ = 1, 2, 3) radii $R_s^{(n)}$ and $R_p^{(n)}$ for three values of the potential well depth. In parentheses, we give the positions of zeroes of the functions $\tau_0(R)$ and $\tau_1(R)$ in Figure 4. All quantities in Table 2 are in atomic units. As it is seen, correlation between the zero positions of the curves on the $R$-axis and the values of $R_l^{(n)}$ is undoubted. Note that while for the functions $\tau_0(U)$ and $\tau_1(U)$ at small electron energies, the critical values of the potential well depths $U^{(i)}$ coincide with zeroes of the functions (Figure 1), for the functions $\tau_0(R)$ and $\tau_1(R)$ we see closeness, but not coincidence between their zeroes and the critical values of $R^{(i)}$ (see Figure 3)

Table 2. Positions of the EWS-time delay zeroes at the $R$ axes
S-wave

| $U$ | $R_s^{(1)}$ | $R_s^{(2)}$ | $R_s^{(3)}$ |
|---|---|---|---|
| 0.05 | 4.967 (4.72) | 14.902 (14.22) | 24.836 (-) |
| 0.1 | 3.512 (3.42) | 10.537 (10.27) | 17.562 (17.15) |
| 0.15 | 2.868 (2.82) | 8.604 (8.45) | 14.339 (14.10) |

P-wave

| $U$ | $R_p^{(1)}$ | $R_p^{(2)}$ | $R_p^{(3)}$ |
|---|---|---|---|
| 0.05 | 9.935 (9.48) | 19.869 (18.96) | 29.804 (-) |
| 0.1 | 7.025 (6.84) | 14.050 (13.72) | 21.074 (-) |
| 0.15 | 5.736 (5.64) | 11.471 (11.28) | 17.207 (16.92) |

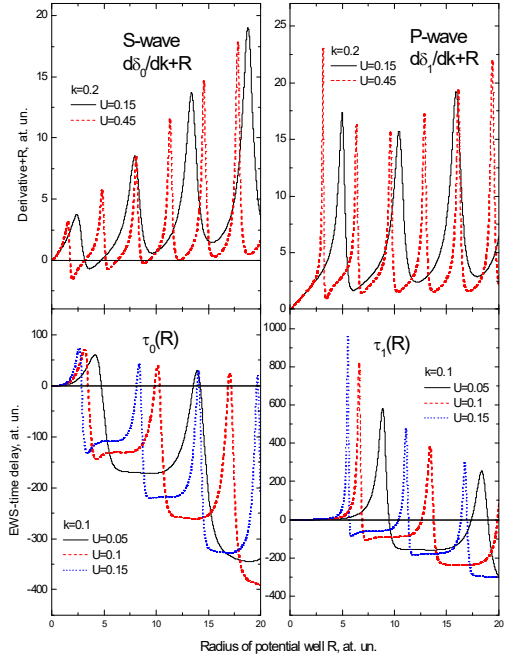

Figure 4. The EWS-times delay $\tau_0(R)$ and $\tau_1(R)$ as functions of $R$ for the fixed electron wave numbers $k$ and the fixed potential depths $U$. The left graphs are for the $s$-partial wave; the right graphs are for the $p$-partial wave. Upper graphs preset the sums $\dot{\delta}_l(R) + R$.

## 5. Conclusions

We have investigated the partial EWS-times delay for slow electrons scattered by rectangular attractive potentials as functions of the potential parameters, such as the potential well depth and the potential radius. We have focused our consideration on the vicinities of the potential parameters that are close to their critical values. The critical values are those, at which bound states with zero binding energy appear in the potential well. Specifically, we have considered potential depths $U$ and potential radii $R$, at which the potential supports some number of discrete $s$-, $p$- and $d$-levels. In spite of the potential simplicity, the presented analysis makes it possible to find some specific features in the time delay behavior that have general character. They are the following:
1) The EWS-times delay of slow electron scattered by a shallow potential well with no bound discrete levels are always positive. They switch the sign when the first discrete level appears regardless of whether this level appears due to changes in the depth of the well *U* or its radius *R* (see Figures 1 and 4).

2) Small changes of the potential well parameters in the vicinities of the first and any other discrete levels arising lead to sudden jumps of the EWS-times delay from a positive value to a negative one. The amplitudes of these jumps increase with decrease of the electron wave number *k* (see Figure 1). With increasing electron momentum *k* (see panels in the second and third rows in figure 1) zeroes of the function $\tau_l(U)$ are slightly shifted from their positions at *k*=0 (see Table 1).

3) The derivatives of the phase shifts of the electron wave functions $\dot{\delta}_l(k)$ are characterized by different asymptotic behavior when *k*→0 (see Figure 2). In the case of *s*-wave, the derivative $\dot{\delta}_0(k)$ approaches finite values different from zero, whereas for waves with $l > 0$ one has $\dot{\delta}_{l>0}(k \to 0) = 0$. When *s*- and *p*-bound states appear in the well the corresponding partial EWS-times delay $\tau_0(k)$ and $\tau_1(k)$ abruptly change their sign (Figure 3).

4) The partial EWS-times delay as functions of the potential well radius depicted in Figure 4 demonstrated the following. With *R* increase and passing the critical values, the electron retention by the potential well (positive EWS-times delay) alters into a mode, at which the incident electron hardly enters the scatterer (negative EWS-times). This behavior of the EWS-times delay is universal since it takes place not only for the first but also for following arising discrete levels.

Let us emphasize that although the investigated features of the EWS-times delay we obtained using a simple square well potential, it is easy to demonstrate the existence of the same features for any short-range potential. Indeed, the general formulas in Appendix including Eq. (A4) do not have any limitations on using different shapes of potentials.

Of interest would be to measure the times delay in electron scattering processes. Currently, the experimental activity in this area concentrates on investigation of the temporal picture of photoionization of atoms and molecules using high intensity attosecond pulses. In this case, the shortness of the light pulse leads to a big spread in the incoming photon energy due to the uncertainty principle. Note, that the results presented in this paper require high energy resolution for the incident electrons, thus imposing strong limitation on the accuracy of time delay measurements.

**Acknowledgments**
ASB is grateful for the support to the Uzbek Foundation Award OT-Ф2-46.

**Appendix A**

The starting point for calculation of the derivatives of the *s*- and *p*-phase shifts in Section 2 is the equality for the logarithmic derivatives of the electronic wave functions at the border of the potential well. Such an approach, being very efficient and simple for *l*=0 and *l*=1, becomes cumbersome for *d*-phase shifts. Below we present another method for the same task.

The radial part of the electron continuum wave function $R_{kl}(r)$ Eqs. (4) and (5) obeys to the following wave equation

$$P_{kl}'' - 2V(r)P_{kl} - \frac{l(l+1)}{r^2}P_{kl} + k^2 P_{kl} = 0, \quad (A1)$$

where $P_{kl}(r) = rR_{kl}(r)$. We assume here, as before, a finite range $R$ of the potential $V(r)$, and we assume the function $V(r)$ to be finite everywhere.

Applying the operator $\partial/\partial k$ to the wave equation (A1), we obtain

$$\dot{P}_{kl}'' - 2V(r)\dot{P}_{kl} + k^2 \dot{P}_{kl} + 2kP_{kl} - \frac{l(l+1)}{r^2}\dot{P}_{kl} = 0. \quad (A2)$$

Multiplying Eqs. (A1) and (A2) by $\dot{P}_{kl}$ and $P_{kl}$, respectively, and subtracting them, we obtain

$$(\dot{P}_{kl}P_{kl}' - P_{kl}\dot{P}_{kl}')' - 2kP_{kl}^2 = 0. \quad (A3)$$

Integrating Eq. (A3) between the radius values 0 and $r$, and using the boundary conditions $P_{kl}(r=0) = 0$, we obtain

$$(\dot{P}_{kl}P_{kl}' - P_{kl}\dot{P}_{kl}') - 2k\int_0^r P_{kl}^2(\rho)d\rho = 0 \quad (A4)$$

This equation does not imply any specific shape of the potential function $V(r)$. Indeed, the first term in this equation refers to the region beyond the potential well ($r \geq R$), and does not depend on the shape of the potential well. For this reason, the first term is the same for any short-range potential function. The specific type of particle interaction with the scattering potential $V(r)$ is totally confined by the the integral in Eq. (A4). Thus, we have two types of terms in (A4). The first of them depends on $k$ and $r$ only while the integral in (A4) is a function of the well parameters.

Let us apply Eq. (A4) to the rectangular potential well Eq. (2). For simplicity, we restrict ourselves by using the $s$-partial wave. In this case for $r \geq R$, we have

$$P_{k0}(r) = \sin(kr + \delta_0) \quad (A5)$$

For distances $r \geq R$, we substitute Eq. (A5) into the first term of Eq. (A4). The result is as follows

$$\dot{\delta}_0 = -r + \frac{1}{2k}\sin 2(kr + \delta_0) + 2\int_0^r P_{k0}^2(\rho)d\rho, \quad (r \geq R). \quad (A6)$$

Note, that this general formula for the s-wave was derived in [18]. The wave function $P_{k0}(r)$ in Eq. (A6) has the form

$$P_{k0}(r) = A_0 \sin qr \qquad (A7)$$

inside the potential well ($r<R$). For $r>R$, the wave function becomes Eq. (A5). Matching the wave functions Eq. (A5) and Eq. (A7) at $r = R$ gives the amplitude $A_0$ in Eq. (A7)

$$A_0 = \frac{\sin(kR+\delta_0)}{\sin qR}. \qquad (A8)$$

In the spherical rectangular potential well, the integral in Eq. (A6) is the sum of two following integrals

$$\int_0^R \sin^2 qr'dr' = \frac{R}{2}\left(1 - \frac{\sin 2qR}{2qR}\right), \qquad (A9)$$

and

$$\int_R^r \sin^2(kr'+\delta_0)dr' = \frac{1}{4k}[2(kr+\delta_0) - \sin 2(kr+\delta_0) - 2(kR+\delta_0) + \sin 2(kR+\delta_0)]. \qquad (A10)$$

Substituting both Eq. (A9) and Eq. (A10) into Eq. (A6), we obtain for the derivative $\dot\delta_0$

$$\dot\delta_0 = -R + \frac{\sin 2(kR+\delta_0)}{2k} + A_0^2 R\left(1 - \frac{\sin 2qR}{2qR}\right). \qquad (A11)$$

As expected, the derived formula (A11) is identical to that of Eq. (12) obtained by differentiation of the boundary condition Eq. (10).
Applying the general formula (A4) to the d-partial wave and dropping down rather cumbersome intermediate computations, we obtain the following expression for the derivative $\dot\delta_2(k)$

$$\dot\delta_2 = -R + \left(1 - \frac{12}{k^2R^2}\right)\frac{\sin 2(kR+\delta_2)}{2k} + \frac{6\sin^2(kR+\delta_2)}{k^4R^3} + \frac{6\cos^2(kR+\delta_2)}{k^2R} + W_2(x,y) \qquad (A12)$$

where the function $W_2(x,y)$ has the form

$$W_2(x,y) = A_2^2 R\left[1 - \frac{3}{x^2} - \frac{3}{x^4} + \left(\frac{6}{x^3} - \frac{1}{2x}\right)\sin 2x + 3\left(\frac{1}{x^4} - \frac{1}{x^2}\right)\cos 2x\right]. \qquad (A13)$$

Here

$$A_2 = \frac{(3-y^2)\sin(y+\delta_2) - 3y\cos(y+\delta_2)}{y^2[(3-x^2)\sin x - 3x\cos x]}, \quad (A14)$$

where the dimensionless variables are $x = qR$ and $y = kR$, as before.

The numerical calculations indicated that the qualitative behavior of the function (A12) versus variables $k$, $R$ and $U$ is similar to that discussed above for $s$- and $p$-waves. The critical values of the first three $d$-levels in the well with $R=2$ are: $U_d^{(n)}$=2.533, 7.508, and 14.857 from Eq. (26).

The following explicit expressions derived from the general formula Eq. (9) determine the $s$-, $p$- and $d$-phase shifts referred to in the text of the paper and in Appendix

$$\tan(y+\delta_0) = \frac{y}{x}\tan x,$$

$$\tan(y+\delta_1) = \frac{x^2 y \tan x}{xy^2 + (x^2 - y^2)\tan x},$$

$$\tan(y+\delta_2) = \frac{3xy(y^2 - x^2) + [3x^2 + y^2(x^2 - 3)]y\tan x}{[y^2(3+y^2) - 3x^2]x + 3(x^2 - y^2)\tan x}. \quad (A15)$$

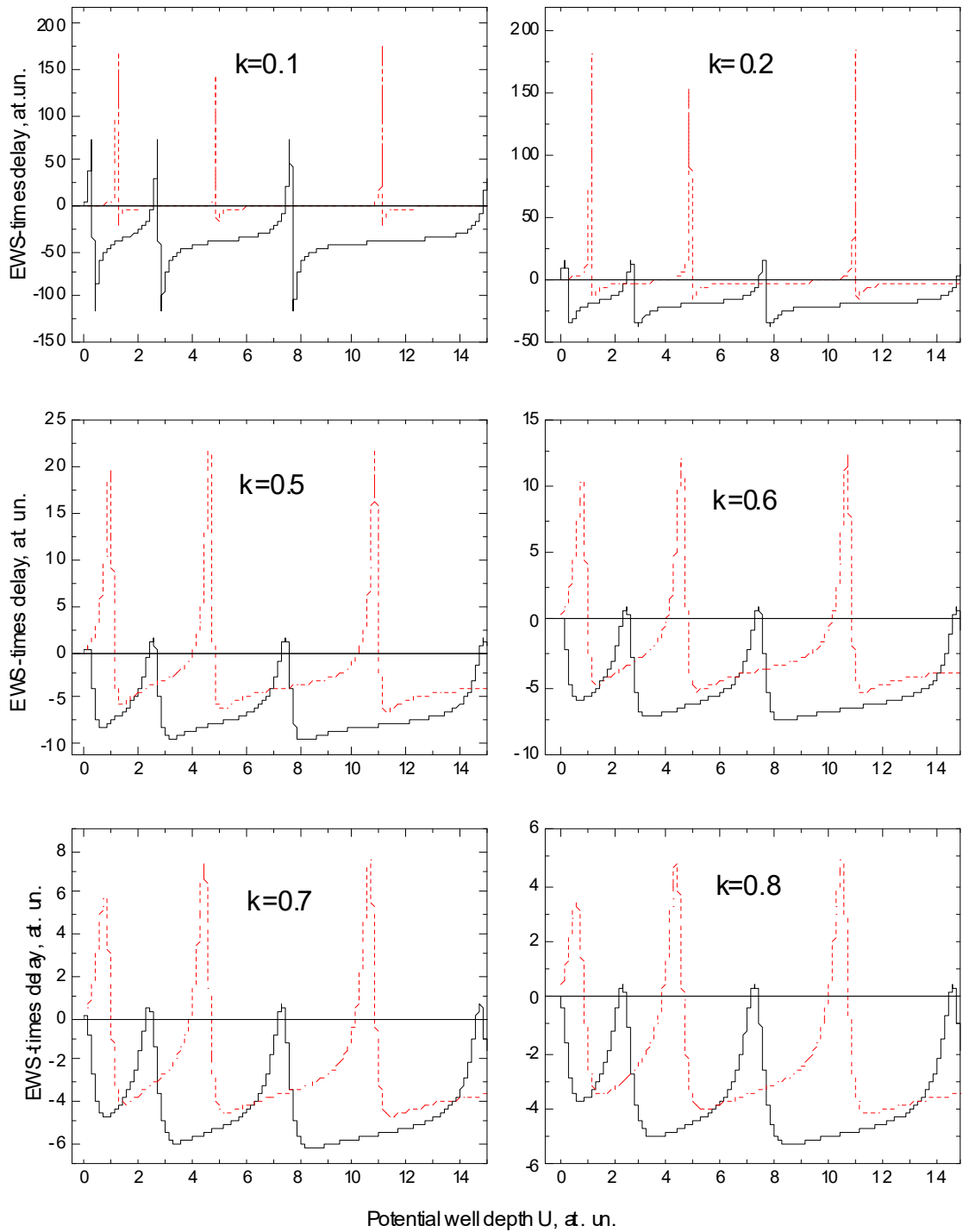

Figure 1. Partial EWS-times delay versus the potential well depths *U* for different electron momenta *k*. Black solid lines are for the *s*-partial wave; dashed red lines are for the *p*-wave.

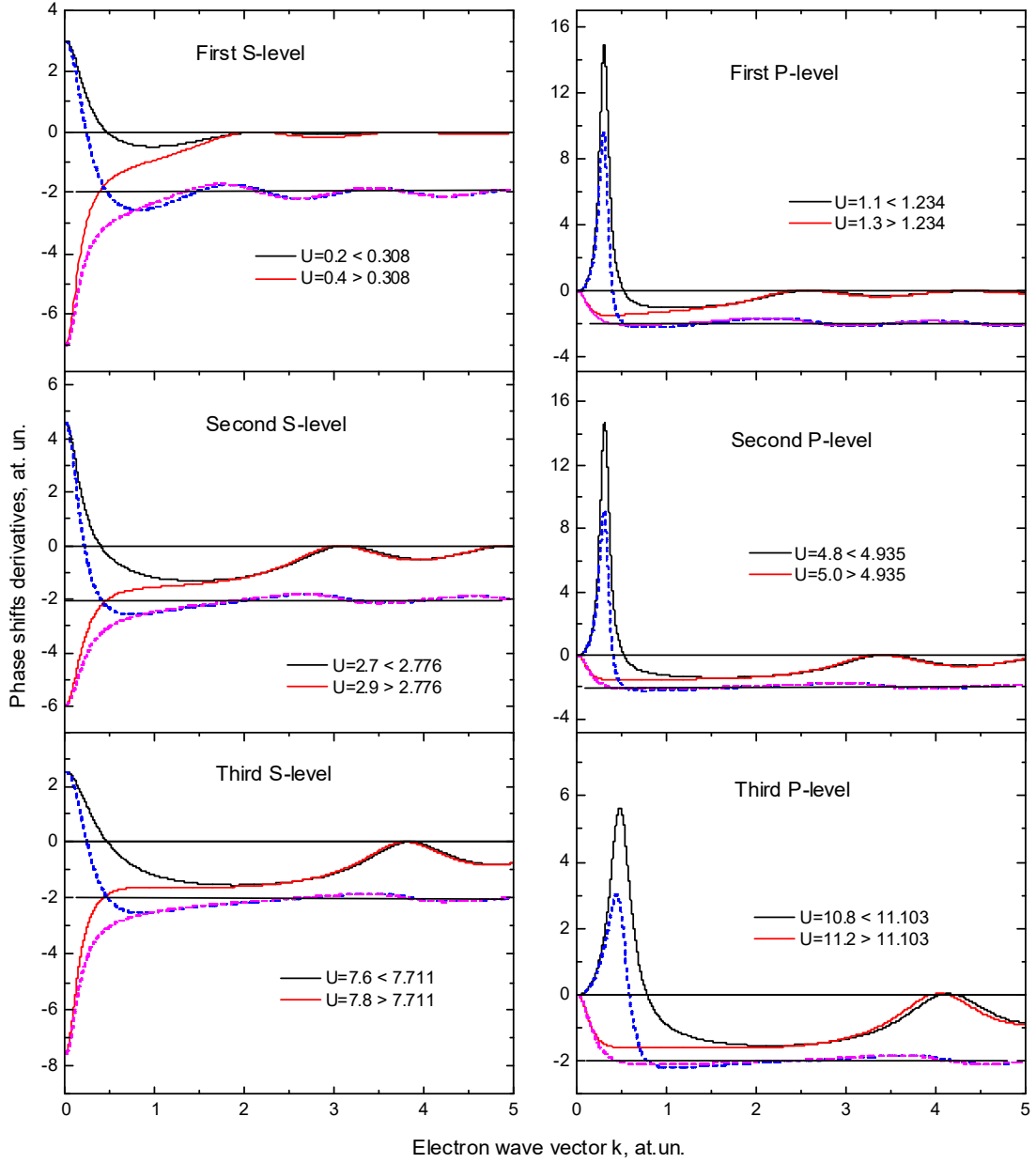

Figure 2. The *s*- and *p*-phase shift derivatives as functions of the electron wave vector *k*. Left panel is the *s*-phase shift; right one is the *p*-phase shift. Solid black and solid red lines are for the phase shifts for potential wells with depths before and after level arising, respectively. The critical values of the first three levels in the wells are: $U_s^{(n)}$=0.308, 2.776, and 7.711 from Eq. (23) for the *s*-state and $U_p^{(n)}$=1.234, 4.935, and 11.103 from Eq. (25) for the *p*-state. Dashed blue and magenta lines present the calculations with the Wignerformulas Eqs. (16) and (17), correspondingly.

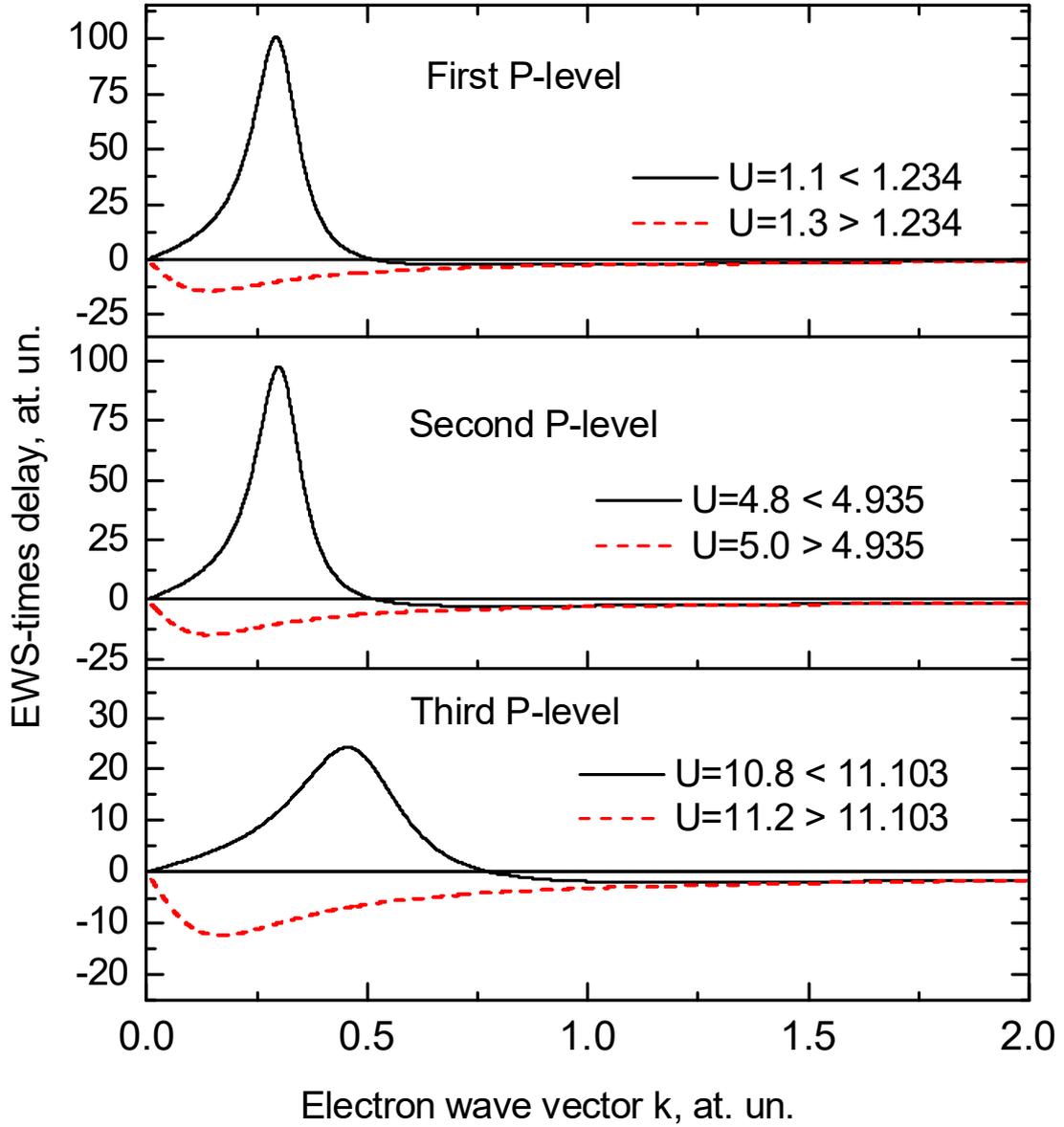

Figure 3. The EWS-times delay $\tau_1(k)$ as functions of the electron wave vector $k$ in the vicinities of the first three *p*-levels arising in the potential well. Solid black lines are for the depth before level arising; dashed red lines are for the depth after *p*-level arising.

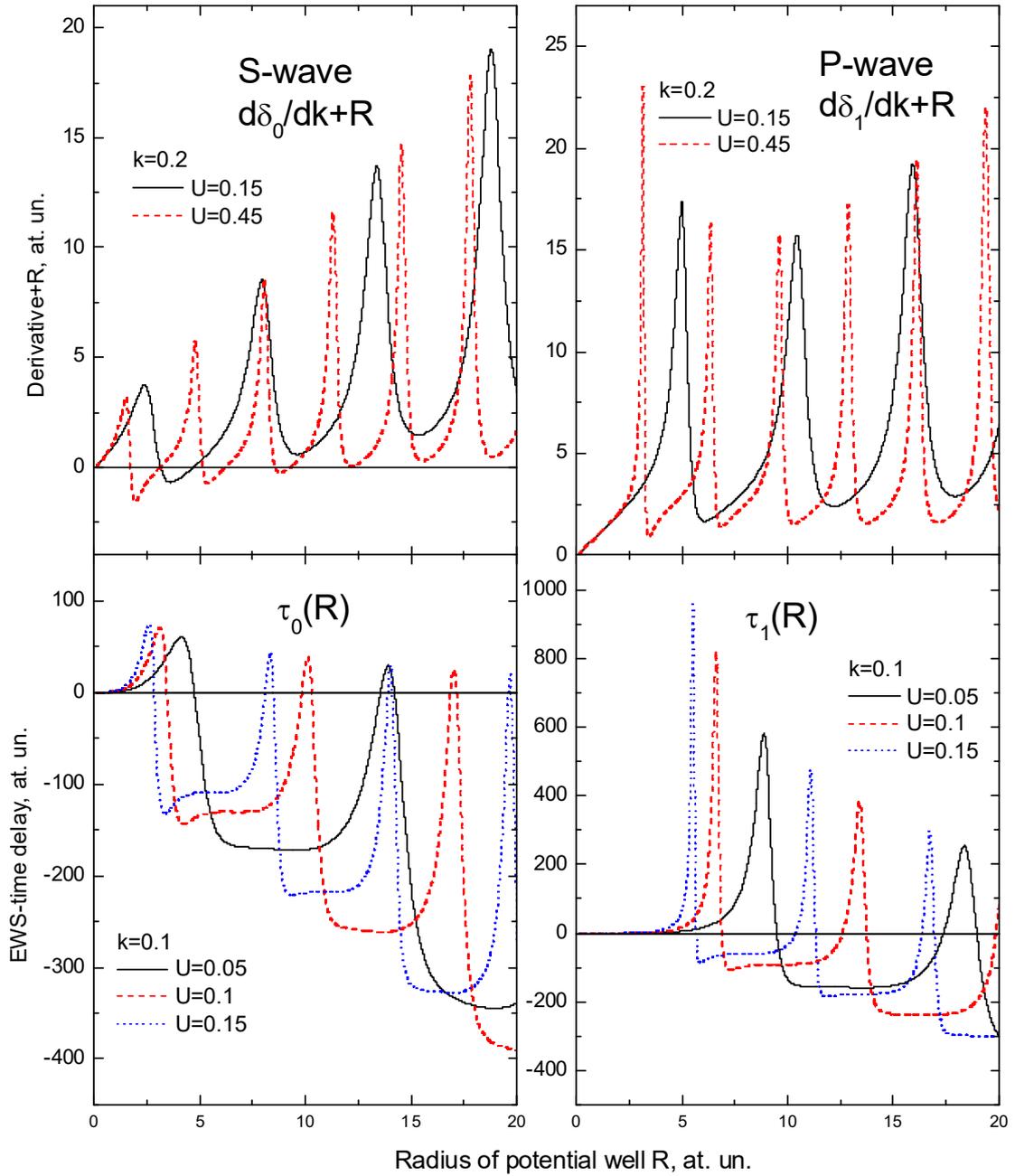

Figure 4. The EWS-times delay $\tau_0(R)$ and $\tau_1(R)$ as functions of $R$ for the fixed electron wave numbers $k$ and the fixed potential depths $U$. The left graphs are for the $s$-partial wave; the right graphs are for the $p$-partial wave. Upper graphs preset the sums $\dot{\delta}_l(R)+R$.